\documentclass[12pt]{article}

\usepackage[totalheight = 23cm, totalwidth = 17cm]{geometry}
\usepackage{amssymb,amsmath,amsfonts,amsbsy,epsfig}

\begin{document}

\title{Generation of perturbation after multi-field inflation}

\author{
Jinn-Ouk Gong\footnote{jgong@hri.res.in}
\\ \\
\textit{Harish-Chandra Research Institute}
\\
\textit{Chhatnag Road, Jhunsi}
\\
\textit{Allahabad 211 019}
\\
\textit{India} }

\date{\today}

\maketitle

\begin{abstract}

We explore a new possibility that some inflaton fields in multi-field inflation
models satisfy the observed value of the spectral index so that the curvature
perturbation generated by them through post-inflationary dynamics may be relevant
for the current observations. We illustrate this point using a simple and
reasonable multi-field chaotic inflation model motivated from string theory.
Possible non-Gaussianity and the tensor-to-scalar ratio are briefly addressed.

\end{abstract}

\thispagestyle{empty} \setcounter{page}{0}
\newpage
\setcounter{page}{1}

Now it is widely accepted that an inflationary epoch at the earliest moment in the
history of the universe \cite{inflation} provides the initial conditions for the
standard hot big bang cosmology and solves many cosmological problems such as the
horizon problem. At the same time, it is usually assumed that the vacuum
fluctuations of the scalar field which gives rise to inflation, the inflaton, are
responsible for the primordial density perturbations \cite{perturbation}. A
particularly convenient quantity to study these perturbations is the curvature
perturbation defined on comoving or uniform energy density hypersurfaces: the former
and the latter are usually written as $\mathcal{R}_c$ and $\zeta$ respectively, and
are equivalent on super-horizon scales.

Quite frequently the models are invoked where inflation is driven by only one
inflaton field as the simplest possibility. Since there exists only a single degree
of freedom during inflation, the resulting curvature perturbation is in nature
adiabatic and practically constant after horizon crossing. This renders the
calculation of the curvature perturbation very simple and systematic, as accurate as
we want \cite{accurate}. However, an immediate difficulty from the particle physics
point of view is that, typically the simplest possibility of chaotic inflation with
power law type potential requires an initial value of the inflaton field far larger
than the Planck scale $m_\mathrm{Pl} \equiv (8\pi G)^{-1/2} \approx 2.4 \times
10^{18} \mathrm{GeV}$. Thus the form of the effective potential is under yet
uncontrollable radiative corrections, quantum gravity effects, and so on. Meanwhile,
in multi-field models all the fields participating in the inflationary dynamics can
have sub-Planckian amplitudes throughout inflation since the Hubble friction
receives contributions from every field. This fact can lead to the desirable
slow-roll phase which is not possible at all for single field case with
sub-Planckian field value, and gives better theoretical control on the models.
Moreover, in theories beyond the standard model of particle physics which might be
the plausible framework for implementing inflation in the early universe, there
exist plenty of scalar fields \cite{bsmreview}. Thus it is natural to consider
multi-field inflation models within such theories \cite{multireview}. In multi-field
inflation models, to calculate the curvature perturbation we can resort to the
so-called $\delta \mathcal{N}$ formalism \cite{deltaN} so that all the light fields
come together to contribute to the final curvature perturbation.

However, the existence of a number of light scalar fields opens many interesting
possibilities for the generation of perturbations other than by the fluctuations of
the inflaton field, e.g. curvaton mechanism\footnote{In the present note, the
curvaton mechanism is supposed to encompass a bit wider context compared with what
is usually mentioned: we mean the curvaton mechanism for the general case where one
or more scalar fields come to dominate the energy density of the universe in the
radiation background due to the oscillation around the minimum of the effective
potential, and thus the curvature perturbations associated with them become
relevant. These fields are not necessarily negligibly light during inflation, which
is the case of the `usual' curvaton scenario.} \cite{curvaton} and inhomogeneous
reheating scenario \cite{inhomogeneousreheating}. Indeed, if the resulting amplitude
of the curvature perturbation generated during inflation is not sufficient to
satisfy the observations, other scenarios would provide the dominant mechanism for
the generation of the curvature perturbation. It is fair to say that this
possibility is open and hence deserves further study. If this is the case, not only
the amplitude, but also the spectral index of the corresponding field fluctuations
should be what we have observed \cite{observation},
\begin{equation}\label{observedindex}
n - 1 \approx -0.052_{-0.018}^{+0.015} \, .
\end{equation}
This is in fact a very strong constraint on the possible mechanism for the
generation of curvature perturbation. In $\delta\mathcal{N}$ formalism, however, the
collective contributions to $n$ is calculated but the index of each inflaton field
is not considered. On the contrary, for the usual curvaton and other scenarios the
relevant fields are completely negligible during inflation. In any case, the
possible contribution of the individual inflaton field after inflation has not yet
been considered. In this note, we consider a new possibility that some of the
inflaton fields may satisfy Eq.~(\ref{observedindex}) so that they are responsible
for the generation of perturbation after inflation.

First let us recall the notion of the `inflaton' field in multi-field inflation
models. During inflation, we may choose a basis in the field space in which the
field evolution is aligned to one field direction $\phi$, or the {\em inflaton}, and
the fluctuations associated with $\phi$ are responsible for the time independent
`adiabatic' component of the perturbations. At a given point where the quantities of
our interest are evaluated, $V$ has the steepest descent along the $\phi$ direction.
This component has the well known result of the spectral index \cite{standardtext},
\begin{equation}
n_\phi - 1 = 2\eta_\phi - 6\epsilon \, ,
\end{equation}
where the slow-roll parameters, evaluated at a particular moment of horizon
crossing, are given by
\begin{equation}
\epsilon \equiv -\frac{\dot{H}}{H^2} \, \hspace{0.5cm} \mbox{and} \hspace{0.5cm}
\eta_\phi \equiv \frac{V_{\phi\phi}}{3H^2} \, .
\end{equation}
Here, the derivative $V_{\phi\phi} \equiv d^2V/d\phi^2$ is taken {\em along} the
field trajectory. The curvature of the potential along the other {\em orthogonal}
directions $\sigma_i$ is smaller than that associated with the $\phi$ direction, so
that the primary evolution is along $\phi$ at that very moment. If some of these
directions in the $(\phi,\sigma_i)$ basis are extremely flat, which is usually taken
to be the case for e.g. the curvaton field, the field trajectory remains always
orthogonal to those $\sigma_i$ directions during inflation. In this case, such
$\sigma_i$ are practically decoupled from the inflationary dynamics except adding
some almost constant vacuum energy during inflation, which is not too large to make
inflation eternal. The fluctuations associated with these `isocurvature' fields
$\sigma_i$ have a little different spectral indices \cite{multireview,deltaN}, which
is again evaluated at a specific moment,
\begin{equation}\label{orthogonalindex}
n_\sigma - 1 = 2\eta_\sigma - 2\epsilon \, .
\end{equation}

Now let us return to the subject of our interest, namely, the question {\em whether
the spectral index of the fluctuations associated with some inflaton fields in
multi-field inflation models can be negative to satisfy Eq.~(\ref{observedindex})}.
Here, the term `inflaton fields' denote the fields with big enough masses to affect
the field trajectory appreciably during inflation. In general, the curvature
associated with $\sigma_i$ directions at the moment when the scales of observational
interest exit the horizon is relatively small, but it will later give rise to a
curved field trajectory. This is clearly different from the previously mentioned
`curvaton type' fields, which practically do not affect the field trajectory during
inflation\footnote{Hence, from now on, we disregard such curvaton type fields among
the orthogonal fields $\sigma_i$.}. Usually $\epsilon \lll 1$ for a wide class of
realistic inflation models when the scales corresponding to the current Hubble
radius exit the horizon. Thus the spectral index of the orthogonal fields, which we
can always find in the field space by appropriate transformation, given by
Eq.~(\ref{orthogonalindex}), receives contributions mostly from $\eta_\sigma$, which
seems to be negative to satisfy Eq.~(\ref{observedindex}). In the following we
examine this point using a simple example. Before then, at this point, we would like
to stress that our main concern here is the spectral index, not the observed
magnitude of the curvature perturbation,
\begin{equation}\label{observedP}
\mathcal{P}^{1/2} \sim 5 \times 10^{-5} \, .
\end{equation}
As stated before, there exist many different ways of generating curvature
perturbation other than the standard one generated by the fluctuations in the
inflaton field. Thus it should not be impossible to match Eq.~(\ref{observedP}) by
appropriate post-inflationary process\footnote{Note that in Ref.~\cite{jgong2007} it
is shown that the amplitude of $\mathcal{P}$ under the potential given by
Eq.~(\ref{potential}) is dependent on the average mass scale $\langle m \rangle$.
Thus, for example if we have $\langle m \rangle \sim \mathcal{O}(\mathrm{TeV})$,
post-inflationary generation of curvature perturbation is indispensable.}: in the
curvaton scenario, for example, the spectrum is known to be \cite{curvatonP}
\begin{equation}
\mathcal{P}^{1/2} = \frac{2}{3}rq\frac{H_\star}{2\pi\sigma_\star} \, ,
\end{equation}
where $r = \rho_\sigma/\rho|_\mathrm{dec}$ is the ratio of the curvaton energy
density to the total energy density of the universe at the epoch of curvaton decay,
$q \lesssim 1$ is a constant, and the subscript $\star$ denotes the moment of
horizon crossing. The spectral index $n$, however, is very hard to get modified once
it is generated, and thus can put very strong constraints on the mechanism of
perturbation generation.

\begin{figure}[h]
\begin{center}
\epsfig{file = 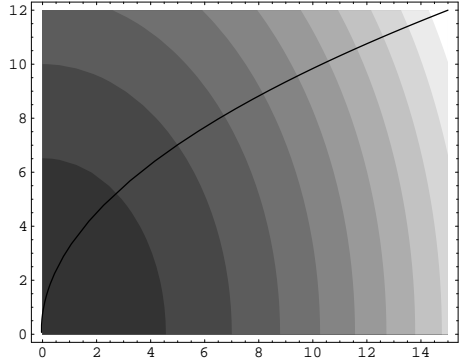, width = 6cm}
\end{center}
\caption{A typical plot of the field trajectory on the two-field potential $V =
m_1^2\phi_1^2/2 + m_2^2\phi_2^2/2$ \cite{doubleinflation}. The horizontal and
vertical axes correspond to $\phi_1$ and $\phi_2$, respectively. Since here we take
$m_1 > m_2 = 0.7 m_1$, the potential is steeper in the $\phi_1$ direction, i.e. the
contour lines are denser along $\phi_1$ direction, and hence the trajectory is along
this direction significantly at the early times. But appreciable $m_2$ makes the
trajectory ``curved'' towards the $\phi_2$ direction later.} \label{2field}
\end{figure}

For definiteness, we consider the simplest case of the multi-field chaotic
inflation,
\begin{equation}\label{potential}
V = \sum_{i=1}^N V_i =  \frac{1}{2} \sum_{i=1}^N m_i^2\phi_i^2 \, ,
\end{equation}
which may be motivated by the ubiquitous string axion fields in a specific string
compactification\footnote{This approximation might be more widely applicable beyond
string theory provided that the fields are displaced not too far from their minima
with their interaction being very weak. For example, for the simplest curvaton
scenario to work properly, it is known that during inflation $H_\star \gtrsim 10^7
\mathrm{GeV}$, or equivalently, $\sigma_\star \gtrsim 10^{-8} m_\mathrm{Pl}$
\cite{curvatoncondition}. As long as the fields of our interest maintain such
amplitude, Eq.~(\ref{potential}) would be a good approximation.} \cite{Nflation}. In
general the masses would not be exactly the same, and without loss of generality we
can place the order of fields by their masses so that $m_1$ corresponds to the most
massive field. The system is described by simple $N + 1$ coupled equations: the
Friedmann equation and the equation of motion of $\phi_i$ are written as
\begin{equation}\label{friedmann}
H^2 = \frac{1}{3m_\mathrm{Pl}^2} \sum_i \left( \frac{1}{2}\dot\phi_i^2 +
\frac{1}{2}m_i^2\phi_i^2 \right) \, ,
\end{equation}
and
\begin{equation}\label{eom}
\ddot\phi_i + 3H\dot\phi_i + m_i^2\phi_i = 0 \, ,
\end{equation}
respectively. In Fig.~\ref{2field} is shown a typical field trajectory in the case
$N = 2$\footnote{Note that exactly the same potential was studied in
Ref.~\cite{bl2002} in the context of the curvaton scenario. However, it is very
important that in general the curvaton should be identified as the `orthogonal'
field which would be given by a linear combination of $\phi$ and $\sigma$, not
simply the lighter one.}.

From Eqs.~(\ref{friedmann}) and (\ref{eom}), the slow-roll parameter $\epsilon$ is
given by \cite{al2005}
\begin{equation}\label{multiepsilon}
\epsilon = \frac{2m_\mathrm{Pl}^2 \sum_i m_i^4\phi_i^2}{\left( \sum_j m_j^2\phi_j^2
\right)^2} \, .
\end{equation}
The other parameters $\eta_i$ are, along each direction,
\begin{equation}\label{multieta}
\eta_i = \frac{2m_\mathrm{Pl}^2m_i^2}{\sum_j m_j^2\phi_j^2} \, .
\end{equation}
First consider the simplest case where all the masses are the same. Then,
Eq.~(\ref{potential}) enjoys perfect rotational symmetry, and this makes every
direction physically indistinguishable and thus becomes practically single field
inflation. Hence we can set the field trajectory a straight line along one
direction, say $\phi_I$, with the other orthogonal fields zero
\cite{al2005,scaleinviso}. Then we find
\begin{equation}
\epsilon = \eta_i = \frac{2m_\mathrm{Pl}^2}{\phi_I^2} \, ,
\end{equation}
so that for the orthogonal fields the spectral index of the field fluctuations,
given by Eq.~(\ref{orthogonalindex}), is almost exactly scale invariant. This is an
interesting but extremely simple limit: generally we should expect that during
inflation the field follows a curved trajectory due to unequal masses.

Now let us break this symmetry in the next simplest two-field case with $m_1 > m_2$
as shown in Fig.~\ref{2field}. Note that the field trajectory along any of these
axes will be a straight line. Thus the two directions $\phi_1$ and $\phi_2$ give the
extrema of the curvature of the potential, with along $\phi_1$ ($\phi_2$)
corresponding to the maximum (minimum). Now let us assume that the field trajectory
is a straight line on the $\phi_1$ axis ($\phi_2 = 0$). Then, the orthogonal
direction to the field trajectory is the $\phi_2$ direction, and correspondingly
\begin{equation}
\eta_\sigma = \eta_2 = \frac{2m_\mathrm{Pl}^2m_2^2}{m_1^2\phi_1^2} = \epsilon
\frac{m_2^2}{m_1^2} \, ,
\end{equation}
and
\begin{equation}
\eta_\phi = \eta_1 = \epsilon = \frac{2m_\mathrm{Pl}^2}{\phi_1^2} \, .
\end{equation}
If the orthogonal fluctuations, currently $\phi_2$, are responsible for the
curvature perturbation we can observe now, then the spectral index, given by
Eq.~(\ref{orthogonalindex}), should be negative, i.e. $\eta_\sigma/\epsilon < 1$. We
can easily see
\begin{equation}\label{eta-epsilonratio}
\frac{\eta_\sigma}{\epsilon} = \frac{\eta_2}{\epsilon} = \frac{m_2^2}{m_1^2} < 1 \,
,
\end{equation}
and indeed in this case the orthogonal field, $\phi_2$, may give dominant
contribution to the curvature perturbation with right spectral tilt due to
post-inflationary dynamics. Note that from Eqs.~(\ref{orthogonalindex}) and
(\ref{eta-epsilonratio}) to match the observations, we need
\begin{equation}
\epsilon \approx \frac{0.025}{1 - m_2^2/m_1^2} \, .
\end{equation}
When the masses are not equal, $\epsilon$ becomes greater compared with that of the
equal mass case and it is unclear by how much it grows without the detail of the
model under consideration \cite{al2005}:  from Eq.~(\ref{potential}), we can find
the number of $e$-folds as
\begin{equation}\label{efolds}
\mathcal{N} \equiv \int H dt = \frac{\sum_i \phi_i^2}{4m_\mathrm{Pl}^2} \, ,
\end{equation}
where we have neglected $\mathcal{O}(1)$ constant. For the equal mass case, it is
equivalent to the single field chaotic inflation. Thus in terms of $\mathcal{N}$, we
can estimate
\begin{equation}
\epsilon > \epsilon_\mathrm{eq} = \frac{1}{2\mathcal{N}} \, .
\end{equation}

It may be noticed that Eq.~(\ref{eta-epsilonratio}) is valid for a special case and
is not applicable in general. However, this can be easily resolved as long as there
are enough number of fields. From Eqs.~(\ref{multiepsilon}) and (\ref{multieta}),
the ratio is written as
\begin{equation}
\frac{\eta_i}{\epsilon} = \frac{m_i^2\sum_j m_j^2\phi_j^2}{\sum_k m_k^4\phi_k^2} \,
.
\end{equation}
Exactly for how many fields this ratio is less than one would depend on the
underlying mass distribution \cite{jgong2007}: for example, as the simplest
possibility assume that the masses are randomly distributed over a range. Then, it
is easy to expect that roughly half of the fields satisfy the condition
$\eta_i/\epsilon < 1$. More importantly, about 50 $\sim$ 60 $e$-folds before the end
of inflation, the field trajectory is almost orthogonal to the most light fields.
Thus, the corresponding $\eta_i$ of these field fluctuations receive contributions
mostly from $m_i$ and the effects by other fields, especially those which provide
the most significant curvature in the field space at that time, are negligible.
However, we stress again that unlike the curvaton type fields the later stage of
inflation is driven by these fields and they are indeed the `inflaton' fields.

Finally, let us briefly examine if this is really the case, what would be the
distinctive observational signatures from the case where the curvature perturbation
is mostly due to the inflaton fluctuations, regarding near future cosmological
observations in particular: non-Gaussianity and the tensor-to-scalar ratio. Denoting
$\mathcal{N}_{,i} \equiv \partial \mathcal{N}/\partial \phi_i$ and
$\mathcal{N}_{,ij} \equiv \partial^2 \mathcal{N}/\partial\phi_i\partial\phi_j$ and
using Eq.~(\ref{efolds}), the non-linear parameter $f_\mathrm{NL}$ is given by
\cite{deltaNnG}
\begin{equation}
-\frac{3}{5}f_\mathrm{NL} \approx \frac{\sum_{i,j} \mathcal{N}_{,i}
\mathcal{N}_{,ij} \mathcal{N}_{,j}}{2\left( \sum_k \mathcal{N}_{,k}^{\,\,2}
\right)^2} = \frac{m_\mathrm{Pl}^2}{\sum_i\phi_i^2} = \frac{1}{4\mathcal{N}} \, ,
\end{equation}
which is far less than 1 and is absolutely not observable as long as relevant
non-Gaussian contribution is generated during inflation. However, it is indeed
possible that significant non-Gaussianity is obtained by the post-inflationary
dynamics: e.g. if the curvaton mechanism is relevant \cite{curvatonP,curvatonnG},
\begin{equation}
f_\mathrm{NL} \sim \frac{5}{4r} \, ,
\end{equation}
i.e. a small fraction of such fields would give rise to large non-Gaussianity, along
with considerable isocurvature perturbation between radiation and matter
\cite{bigiso}. It seems impossible to make any completely general and model
independent prediction on the level of non-Gaussianity. Meanwhile, the well known
result for the tensor-to-scalar ratio, $r = 16\epsilon$, gives the ratio of the
tensor perturbations to the scalar perturbations due to the inflaton contribution
only. In the cases of our interest, however, for the scalar perturbations there
exist contributions other than the inflaton component. Thus we can write
\cite{otherr}
\begin{equation}
r = \frac{\mathcal{P}_\mathrm{T}}{\mathcal{P}_\mathrm{s}} = 16\epsilon
\frac{\mathcal{P}_\phi}{\mathcal{P}_\phi + \mathcal{P}_\sigma} \, ,
\end{equation}
which is always smaller than $16\epsilon$, and can be absolutely negligible if the
inflaton contribution is sub-dominant.

In summary, we have explored the possibility that during multi-field inflation the
spectral indices of the fluctuations of the isocurvature components of the inflaton
field are negative. Indeed, at least for some of them the indices are negative and
may match the observed value. Therefore as far as the spectral index is concerned,
they can be plausible candidates for curvaton type mechanism, and may serve as the
dominant contributions to the curvature perturbation relevant for our observable
universe, depending on the post-inflationary dynamics. This suggests that the
generation of the curvature perturbation in multi-field inflation models is a much
more generic phenomenon than usually perceived, and that the studies on this topic
should be very careful. Concerned with near future observations, compared with the
case where the inflaton contribution is dominant, we cannot make any model
independent prediction on non-Gaussianity. The tensor-to-scalar ratio, however, will
be smaller and may not be detected in any foreseeable observations.

\subsection*{Acknowledgements}

It is a pleasure to thank Ki-Young Choi for useful conversations at the early stages
of this work: the main idea was motivated during a collaboration with him. I am
especially deeply indebted to David Lyth for invaluable comments, discussions and
suggestions at every stage of this work.

\end{document}